# Coupled Rendezvous and Docking Maneuver control of satellite using Reinforcement learning-based Adaptive Fixed-Time Sliding Mode Controller

**Rakesh Kumar Sahoo[a*], Manoranjan Sinha[b]**

rakesh.sahoo266@gmail.com[a], masinha@aero.iitkgp.ac.in[b]

**Department of Aerospace Engineering, Indian Institute of Technology Kharagpur, West Bengal, 721302, India**

***Corresponding author**

## Abstract

Satellite dynamics in unknown environments are inherently uncertain due to factors such as varying gravitational fields, atmospheric drag, and unpredictable interactions with space debris or other celestial bodies. Traditional sliding mode controllers with fixed parameters often struggle to maintain optimal performance under these fluctuating conditions. Therefore, an adaptive controller is essential to address these challenges by continuously tuning its gains in real-time. In this paper, we have tuned the slopes of the Fixed-time Sliding surface adaptively using reinforcement learning for coupled rendezvous and docking maneuver of chaser satellite with the target satellite in an unknown space environment. The neural network model is used to determine the optimal gains of reaching law of the fixed-time sliding surface. We have assumed that we don't have an accurate model of the system so we have added noise in the tangent space instead of directly on the manifold to preserve the geometric structure of the system while ensuring mathematically consistent uncertainty propagation. The reinforcement learning is used as an approximator to represent the value function of the agent to estimate the dynamical model of the system using the Actor-Critic method. The proposed control algorithm integrates a neural network and a sliding mode controller in a cascade loop architecture, where the tracking error dynamically tunes the sliding surface gains. Global fixed-time stability of the closed-loop feedback system is proved within the Lyapunov framework. This comprehensive approach of fixed-time sliding mode controller using a Reinforcement Learning based ensures the completion of the mission efficiently while addressing the critical challenges posed by the uncertain environment. The simulation results presented support the claims made.

## 1 Introduction

The capture of inactive target satellite is a critical challenge in the context of satellite servicing, debris removal, and interplanetary exploration for space sustainability. This capturing process involves rendezvous and docking maneuver (RVD) where the chaser satellite autonomously approach and securely docks with the target satellite [1, 2]. The RVD process involves the precise coordination of the chaser satellite's motion to approach, align, and physically connect with the target satellite [3, 4]. However, a key challenge in RVD is the need for coupled translational and rotational maneuvers [5]. These two types of motion are inherently interdependent [6]. As the chaser satellite approaches the target, it must continuously adjust its orientation to ensure that docking interfaces are accurately aligned for relative navigation, while simultaneously controlling relative velocity to avoid impact during docking [7, 8].

Traditional methods often rely on Euclidean space representations for modelling the translational and rotational dynamics of satellite separately and thus does not account for non-linear coupling between these motions [9, 10]. In contrast, the coordinate-free representation of the Lie group SE(3) captures both translational and rotational dynamics in a unified framework [11, 12]. Lie group, SE(3) combines the special orthogonal group $SO(3)$ for rotations and $\mathbb{R}^3$ for translations, thus avoids singularities [13] and unwinding [14] that arise when using attitude parameterizations such as euler angles, quaternion [15] and modified Rodrigues parameters [16]. The geometric structure of $SE(3)$ facilitates the development of a globally stable controller that are more efficient and easier to implement that the Euclidean counterpart [17]. In the near Earth orbit, the dynamics of the chaser satellite can be modeled accurately, due to the availability of well-defined gravitational models and disturbance parameters. However, challenges arise during deep space exploration, where multiple celestial bodies influence the gravitational forces and gravity gradient torque acting on the chaser satellite resulting in inaccuracies during the rendezvous and docking phase. The inability to accurately model these gravitational effects can lead to substantial errors in state estimation and trajectory predictions, which are critical for successful rendezvous and docking maneuvers. As a result, developing robust algorithms and control strategies that can adapt to these uncertainties is crucial for ensuring successful docking operations.

Sliding Mode Controller (SMC) is a widely used non-linear controller for its robustness and capability to handle non-linearity and disturbances. However, SMC is sensitive to parameter variations and uncertainties [18] as the system states are unconstrained during the reaching phase. To address this, several reaching laws [19] have been proposed in the literature, including the exponential reaching law [20] and signum function-based reaching law [21] for exponential and constant velocity-based convergence rate of the system, respectively. The second approach explored in the literature involves minimizing the reaching phase of the sliding mode controller by employing a time-varying slope of the sliding surface [22, 23, 24]. A third approach leverages optimization algorithms such as particle swarm optimizer [25, 26], genetic algorithm [27, 28], and fuzzy logic-based optimizer [29] to determine the optimal gains of the sliding surface. However, all the above methods for determining the optimal gains of the sliding surface are fundamentally model-based approaches, as they rely on the accurate mathematical model of the system to predict behavior and optimize control inputs. To address these limitations, model-free techniques such as neural networks have emerged as a powerful tool for adaptive tuning of the gains in sliding mode controllers. The integration of neural networks with SMC enables the system to adaptively learn the required gain adjustments to optimize control performance without manual intervention [30, 31]. Moreover, neural network based sliding mode controllers enhance convergence rate and reduce chattering phenomena by estimating uncertainties in real time [32].

In this paper, we have tuned the gains of reaching law of the Fixed-time Sliding surface adaptively using model-free neural network for coupled rendezvous and docking maneuver of chaser satellite with the target satellite in presence of uncertainty. We have added multi-variate Gaussian noise in the tangent space instead of directly on the manifold to preserve the geometric structure of the system while ensuring mathematically consistent uncertainty propagation. Reinforcement learning based on Actor-Critic method is used as an approximator to estimate the dynamical model of the system. The proposed control algorithm integrates a neural network, sliding mode controller and reinforcement learning network in a cascade layer architecture. Global fixed-time stability of the closed-loop feedback system is proved within the Lyapunov framework.

The structure of this paper is as follows: Section 2 introduces the rigid body dynamic model of the target and chaser satellites, followed by relative coupled translational and rotational dynamics. Section 3 outlines the assumptions, definitions, lemmas used, and the hybrid approach of fixed-time sliding mode controller and reinforcement learning techniques. Then the system dynamics of the proposed adaptive controller are analysed in section 4. Finally, the simulation results are presented to substantiate the claims made.

# 2 Rigid Body Dynamic Model of Satellites

This section presents the rigid body kinematics and dynamic model of the target and chaser satellite orbiting Earth under the influence of a central gravitational field. To describe the relative translational and rotational motion between the satellites-three reference frames are defined as the Earth-Centered Inertial frame $E_i$, denoted by $(E_{ix}\ E_{iy}\ E_{iz})$; the target satellite's body-fixed frame $E_t$, denoted by $(E_{tx}\ E_{ty}\ E_{tz})$; and the chaser satellite's body-fixed frame $E_c$, denoted by $(E_{cx}\ E_{cy}\ E_{cz})$.

## 2.1 Kinematics and Dynamics of Target Satellite

The target satellite is modelled as a rigid body orbiting earth whose configuration space is three-dimensional space mathematically modelled using special Euclidean group $SE(3)$. Lie group $SE(3)$ captures both translational and rotational motion which is represented as a semi-direct product of $SO(3) \ltimes \mathbb{R}^3$. Let the position vector of target satellite be denoted by $\tilde{p}_t \in \mathbb{R}^3$ in the Earth-Centered Inertial frame $E_i$ and the attitude of the target satellite be represented by rotation matrix $\boldsymbol{R}_{t/i}$ which transforms from target's body-fixed frame $E_t$ to Earth-Centered Inertial frame $E_i$. The velocity and angular velocity of target satellite in it's own body frame is represented by $\tilde{v}_t \in \mathbb{R}^3$ and $\tilde{\Omega}_t \in \mathbb{R}^3$, respectively. The translational and rotational kinematics of the target satellite can be described as

$$\dot{\tilde{p}}_t = \boldsymbol{R}_{t/i}\tilde{v}_t \quad , \quad \dot{\boldsymbol{R}}_{t/i} = \boldsymbol{R}_{t/i}\boldsymbol{S}(\tilde{\Omega}_t) \tag{1}$$

where $\boldsymbol{S}(\tilde{\Omega}_t)$ is skew-symmetric matrix form of target's angular velocity vector $\tilde{\Omega}_t = [\Omega_{tx} \quad \Omega_{ty} \quad \Omega_{tz}]^T \in \mathbb{R}^3$ defined as (the exponential mapping operation $(.)^{\times}: \mathbb{R}^3 \rightarrow so(3)$ as shown below:

$$\boldsymbol{S}(\tilde{\Omega}_t) = \begin{bmatrix} 0 & -\Omega_{tz} & \Omega_{ty} \\ \Omega_{tz} & 0 & -\Omega_{tx} \\ -\Omega_{ty} & \Omega_{tx} & 0 \end{bmatrix}$$

The translational and rotational dynamics of the target satellite described in its own body-fixed frame can be described as

$$\begin{aligned} m_t\dot{\tilde{v}}_t &= m_t\boldsymbol{S}(\tilde{v}_t)\tilde{\Omega}_t + \tilde{F}_{g/t} + \tilde{F}_{d/t} \\ \boldsymbol{J}_t\dot{\tilde{\Omega}}_t &= \boldsymbol{J}_t\boldsymbol{S}(\tilde{\Omega}_t)\tilde{\Omega}_t + \tilde{M}_{g/t} + \tilde{M}_{d/t} \end{aligned} \tag{2}$$

where $\tilde{F}_{g/t}$ and $\tilde{F}_{d/t}$ are gravity force and disturbance force acting in the target's body-fixed frame, respectively. Similarly, $\widetilde{M}_{g/t}$ and $\widetilde{M}_{d/t}$ are gravity gradient torque and disturbance torque acting in the target's body-fixed frame, respectively. The gravity force including the $J_2$ component, and gravity gradient torque is given by

$$\begin{aligned} \tilde{F}_{g/t} &= -\frac{\mu m_t}{p_t^3}\left(\boldsymbol{R}_{t/i}^T \tilde{p}_t\right) + m_t \boldsymbol{R}_{t/i}^T \tilde{a}_{j_2} \\ \widetilde{M}_{g/t} &= \frac{3\mu}{p_t^5}\left(\left(\boldsymbol{R}_{t/i}^T \tilde{p}_t\right)^{\times} \boldsymbol{J}_t \left(\boldsymbol{R}_{t/i}^T \tilde{p}_t\right)\right) \end{aligned} \tag{3}$$

where $\mu$ is gravitational constant and $\tilde{a}_{j_2}$ is acceleration due to $J_2$ perturbation effects caused by Earth's oblateness. The configuration of target satellite in inertial frame can be represented using special Euclidean group $SE(3)$ as shown below:

$$\boldsymbol{C}_t = \begin{bmatrix} \boldsymbol{R}_{t/i} & \tilde{p}_t \\ 0_{1\times 3} & 1 \end{bmatrix} \in SE(3) \tag{4}$$

The unified velocity vector of the angular velocity and translational velocity of the target satellite can be represented as

$$\tilde{\xi}_t = \begin{bmatrix} \widetilde{\Omega}_t \\ \tilde{v}_t \end{bmatrix} \in \mathbb{R}^6$$

The kinematics of the target satellite described in Eq. 1 can represented in a compact form as

$$\dot{\boldsymbol{C}}_t = \boldsymbol{C}_t \tilde{\xi}_t^{\vee}$$

where $\tilde{\xi}_t^{\vee}$ is the exponential mapping of $\tilde{\xi}_t : \mathbb{R}^6 \rightarrow se(3)$ and is given by

$$\tilde{\xi}_t^{\vee} = \begin{bmatrix} \boldsymbol{S}(\widetilde{\Omega}_t) & \tilde{v}_t \\ 0_{1\times 3} & 0 \end{bmatrix} \in \mathrm{se}(3)$$

Let the unified matrix of mass and inertia of target satellite be represented by

$$\mathbb{I}_t = \begin{bmatrix} \boldsymbol{J}_t & 0_{3\times 3} \\ 0_{3\times 3} & m_t \boldsymbol{I}_3 \end{bmatrix}$$

where $\boldsymbol{I}_3$ is a three-dimensional identity matrix. Let the unified vector of gravitational force and gravity gradient torque, and the unified vector of external distrurbance force and torque acting on the target satellite be represented by $\tilde{\varphi}_{g/t}$ and $\tilde{\varphi}_{d/t}$, respectively, as shown in Eq. 5.

$$\tilde{\varphi}_{g/t} = \begin{bmatrix} \widetilde{M}_{g/t} \\ \tilde{F}_{g/t} \end{bmatrix} \quad , \quad \tilde{\varphi}_{d/t} = \begin{bmatrix} \widetilde{M}_{d/t} \\ \tilde{F}_{d/t} \end{bmatrix} \tag{5}$$

The translational and rotational dynamics of the target satellite described in Eq. 2 can be represented in a compact form as

$$\mathbb{I}_t \dot{\tilde{\xi}}_t = ad^*_{\tilde{\xi}_t} \mathbb{I}_t \tilde{\xi}_t + \tilde{\varphi}_{g/t} + \tilde{\varphi}_{d/t} \tag{6}$$

where $ad^*_{\tilde{\xi}_t}$ is the coadjoint operation of Lie algebra.

## 2.2 Kinematics and Dynamics of Chaser Satellite

The chaser satellite is modelled as a rigid body orbiting earth whose configuration space is Lie group $SE(3)$. Let the position vector of chaser satellite be denoted by $\tilde{p}_c \in \mathbb{R}^3$ in the Earth-Centered Inertial frame $E_i$ and the attitude of the chaser satellite be represented by rotation matrix $\boldsymbol{R}_{c/i}$ which transforms from chaser's body-fixed frame $E_c$ to Earth-Centered Inertial frame $E_i$. The velocity and angular velocity of chaser satellite in it's own body frame is represented by $\tilde{v}_c \in \mathbb{R}^3$ and $\tilde{\Omega}_c \in \mathbb{R}^3$, respectively. The translational and rotational kinematics of the chaser satellite can be described as

$$\begin{aligned} \dot{\tilde{p}}_c &= \boldsymbol{R}_{c/i}\tilde{v}_c \\ \dot{\boldsymbol{R}}_{c/i} &= \boldsymbol{R}_{c/i}\boldsymbol{S}(\tilde{\Omega}_c) \end{aligned} \tag{7}$$

where $\boldsymbol{S}(\tilde{\Omega}_c)$ is the skew-symmetric matrix form of angular velocity vector $\tilde{\Omega}_c = [\Omega_{cx} \quad \Omega_{cy} \quad \Omega_{cz}]^T \in \mathbb{R}^3$ as described above. The translational and rotational dynamics of the chaser satellite described in its own body frame can be written as

$$\begin{aligned} m_c\dot{\tilde{v}}_c &= m_c\boldsymbol{S}(\tilde{v}_c)\tilde{\Omega}_c + \tilde{F}_{g/c} + \tilde{F}_{c/c} + \tilde{F}_{d/c} \\ \boldsymbol{J}_c\dot{\tilde{\Omega}}_c &= \boldsymbol{J}_c\boldsymbol{S}(\tilde{\Omega}_c)\tilde{\Omega}_c + \tilde{M}_{g/c} + \tilde{M}_{c/c} + \tilde{M}_{d/c} \end{aligned} \tag{8}$$

where $\tilde{F}_{g/c}$, $\tilde{F}_{c/c}$ and $\tilde{F}_{d/c}$ are gravity force, external control force and disturbance force acting in the chaser's body-fixed frame, respectively. Similarly, $\tilde{M}_{g/c}$, $\tilde{M}_{c/c}$ and $\tilde{M}_{d/c}$ are gravity gradient torque, external control torque and disturbance torque acting in the chaser's body-fixed frame, respectively. The gravity force including the $J_2$ component, and gravity gradient torque acting in the chaser satellite body-fixed frame has the same form as that of the target satellite as shown in Eq. 3 The configuration of chaser satellite in inertial frame can be represented using special Euclidean group $SE(3)$ as shown below:

$$\boldsymbol{C}_c = \begin{bmatrix} \boldsymbol{R}_{c/i} & \tilde{p}_c \\ 0_{1\times 3} & 1 \end{bmatrix} \in SE(3)$$

The unified velocity vector of the angular velocity and translational velocity of chaser satellite can be represented as

$$\tilde{\xi}_c = \begin{bmatrix} \tilde{\Omega}_c \\ \tilde{v}_c \end{bmatrix} \in \mathbb{R}^6$$

The kinematics of the chaser satellite described in Eq. 7 can represented in a compact form as

$$\dot{\boldsymbol{C}}_c = \boldsymbol{C}_c\tilde{\xi}_c^{\vee} \tag{9}$$

where $\tilde{\xi}_c^{\vee}$ is the exponential mapping of $\tilde{\xi}_c : \mathbb{R}^6 \rightarrow se(3)$ and is given by

$$\tilde{\xi}_c^{\vee} = \begin{bmatrix} \boldsymbol{S}(\tilde{\Omega}_c) & \tilde{v}_c \\ 0_{1\times 3} & 0 \end{bmatrix} \in \mathrm{se}(3)$$

Let the unified matrix of mass and inertia of the chaser satellite be represented by

$$\mathbb{I}_c = \begin{bmatrix} \boldsymbol{J}_c & 0_{3\times 3} \\ 0_{3\times 3} & m_c\boldsymbol{I}_3 \end{bmatrix}$$

where $\boldsymbol{I}_3$ is a three-dimensional identity matrix. Let the unified vector of gravitational force and gravity gradient torque, and the unified vector of external distrurbance force and torque acting on the chaser satellite be represented by $\tilde{\varphi}_{g/c}$ and $\tilde{\varphi}_{d/c}$, respectively, as shown below.

$$\tilde{\varphi}_{g/c} = \begin{bmatrix} \tilde{M}_{g/c} \\ \tilde{F}_{g/c} \end{bmatrix} \quad , \quad \tilde{\varphi}_{d/c} = \begin{bmatrix} \tilde{M}_{d/c} \\ \tilde{F}_{d/c} \end{bmatrix}$$

The translational and rotational dynamics of the chaser satellite described in Eq. 8 can be represented in a compact form as

$$\mathbb{I}_c\dot{\tilde{\xi}}_c = ad^*_{\tilde{\xi}_c}\mathbb{I}_c\tilde{\xi}_c + \tilde{\varphi}_{c/c} + \tilde{\varphi}_{g/c} + \tilde{\varphi}_{d/c} \tag{10}$$

where $\varphi_{c/c}$ is the unified vector of external control torque and force acting on the chaser satellite body-fixed frame as shown in Eq. 11 to track the desired maneuver

$$\tilde{\varphi}_{c/c} = \begin{bmatrix} \tilde{M}_{c/c} \\ \tilde{F}_{c/c} \end{bmatrix} \tag{11}$$

## 2.3 Relative Kinematics and Dynamics of Satellites

Let the desired configuration of chaser satellite with respect to target satellite at the end of docking maneuver be represented by, $\boldsymbol{D}_{c/t} \in SE(3)$ consisting of desired rotation matrix $\boldsymbol{R}_{c/t} \in SO(3)$, which transforms from chaser's body-fixed frame to target's body-fixed frame and $\tilde{p}_{c/t} \in \mathbb{R}^3$ is the desired position of chaser satellite with respect to the target satellite as shown in Eq. 12

$$\boldsymbol{D}_{c/t} = \begin{bmatrix} \boldsymbol{R}_{c/t} & \tilde{p}_{c/t} \\ 0_{1\times 3} & 1 \end{bmatrix} \in SE(3) \tag{12}$$

The current configuration of the chaser satellite $\boldsymbol{C}_c$ in the inertial frame $E_i$ can be obtained using the Eq. 9. The relative configuration $\boldsymbol{H} \in SE(3)$ of chaser satellite with respect to target satellite in the inertial frame can be written as $\boldsymbol{H} = \boldsymbol{C}_t\boldsymbol{D}_{c/t}$. The relative configuration tracking error vector $\tilde{\rho} = [\tilde{\beta} \quad \tilde{\delta}]^T \in \mathbb{R}^6$ consisting of orientation tracking error $\tilde{\beta}$ and position racking error $\tilde{\delta}$, is obtained by performing logarithmic mapping from Lie group to Lie algebra : $SE(3) \rightarrow se(3)$ as shown below.

$$\tilde{\rho}^{\vee} = logm(\boldsymbol{D}_{c/t}^{-1} * H) \tag{13}$$

The relative velocity of the chaser satellite with respect to the target satellite in the chaser's body-fixed frame is expressed as $\tilde{\sigma} = \tilde{\xi}_c - Ad_{\boldsymbol{H}^{-1}}\tilde{\xi}_t$ where $Ad_{\boldsymbol{H}^{-1}}$ is the adjoint operation of Lie group $\boldsymbol{H}^{-1} \in SE(3)$. The relative acceleration of chaser satellite with respect to the target satellite is expressed as

$$\dot{\tilde{\sigma}} = \dot{\tilde{\xi}}_c + ad_{\tilde{\sigma}}Ad_{\boldsymbol{H}^{-1}}\tilde{\xi}_t - Ad_{\boldsymbol{H}^{-1}}\dot{\tilde{\xi}}_t$$

The relative dynamics of chaser satellite with respect to the target satellite is expressed using Eq. 10 into Eq. 14 as

$$\mathbb{I}_c\dot{\tilde{\sigma}} = \tilde{\varphi}_{g/c} + \tilde{\varphi}_{d/c} + \tilde{\varphi}_{c/c} + \mathbb{I}_c(ad^*_{\tilde{\xi}_c}\mathbb{I}_c\tilde{\xi}_c + ad_{\tilde{\sigma}}Ad_{\boldsymbol{H}^{-1}}\tilde{\xi}_t - Ad_{\boldsymbol{H}^{-1}}\dot{\tilde{\xi}}_t) \tag{15}$$

The time derivative of configuration tracking error $\tilde{\rho}$ is given by $\dot{\tilde{\rho}} = \boldsymbol{G}(\tilde{\rho})\tilde{\sigma}$ [33]. The relative dynamics of coupled translational and rotational maneuver described in Eq. 15 along with known initial states $\tilde{\sigma}(t_0)$ at time $t = t_0$ is used to generate relative states of chaser satellite with respect to the target satellite at time $t \geq t_0$.

# 3 Design approach of adaptive fixed-time sliding mode controller

In this section, an adaptive fixed-time sliding mode controller is designed for the rigid body model of satellites modeled in three-dimensional configuration space $SE(3)$. For stability analysis of fixed-time sliding mode controller, assumption, definition and lemmas are given below.

**Assumption 1** The external disturbance force and torque acting on each satellite $\tilde{\varphi}_d \in \mathbb{R}^6$ i.e., solar radiation force and torque, third-body gravitational force and torque are assumed to be bounded $\tilde{\varphi}_{d,s} = \left[\tilde{M}_{d,s} \quad \tilde{F}_{d,s}\right]^T$ such that, $(\tilde{\varphi}_d)_j \leq \left(\tilde{\varphi}_{d,s}\right)_j \; \forall \; j = 1,2 \ldots 6$.

**Definition 1** For vector $\tilde{x} \in \mathbb{R}^n$ and $l \in \mathbb{R}$, the signum function $sig(\tilde{x})$ and its derivative are defined as

$$\begin{aligned} sig^l(\tilde{x}) &= [sign(x_1)|x_1|^l \ldots sign(x_n)|x_n|^l] \\ \frac{d}{dt} sig^l(\tilde{x}) &= l * diag(sign(\tilde{x})^{l-1})\dot{\tilde{x}} \\ \frac{d}{dt}|\tilde{x}|^l &= l * diag(|\tilde{x}|^{l-1})\dot{\tilde{x}} \end{aligned}$$

**Lemma 1** *(Fixed-Time Stability)* [34] *For a nonlinear system expressed as* $\dot{\tilde{x}} = g(\tilde{x})$, $\tilde{x}(0) = \tilde{x}_0, \tilde{x} \in \mathbb{R}^n$ *if there exist a continuously differentiable Lyapunov function* $V(x)$ *in the neighborhood* $N \subset \mathbb{R}^n$ *of origin, which satisfies*

$$\dot{V}(\tilde{x}) \leq \; k_1 V^{l_1}(\tilde{x}) + k_2 V^{l_2}(\tilde{x})$$

*where* $k_1, k_2 > 0$, $0 < l_1 \leq 1$ *and* $l_2 > 1$. *Then the origin of the system is fixed-time stable and the settling time* $T(x)$ *is bounded by:*

$$T(x) \leq T_{max} = \frac{1}{k_1(1 - l_1)} + \frac{1}{k_2(l_2 - 1)}$$

**Lemma 2** *For any vector* $\tilde{x} \in \mathbb{R}^n$, $0 < l_1 \leq 1$ *and* $l_2 > 1$, *the following inequalities hold*

$$\left(\sum_{i=1}^{n} |x_i|\right)^{l_1} \leq \sum_{i=1}^{n} |x_i|^{l_1} \quad , \quad \left(\sum_{i=1}^{n} |x_i|\right)^{l_2} \leq n^{l_2 - 1} \sum_{i=1}^{n} |x_i|^{l_2}$$

## 3.1 Fixed-time sliding surface

The fixed-time sliding surface consisting of the exponential coordinate vectors, $\tilde{\rho}$ and relative velocity, $\tilde{\sigma}$ to track the desired trajectory within fixed time is defined as

$$\tilde{s} \;=\; \tilde{\sigma} + \mathrm{k}_1 sig^{l_1}(\tilde{\rho}) + k_2 sig^{l_2}(\tilde{\rho}) \tag{16}$$

where $\mathrm{k}_1, \mathrm{k}_2 > 0$, $0 < l_1 \leq 1$, $\mathrm{l}_2 > 1$, $\tilde{s} = [s_1, s_2, \ldots, s_6] \in \mathbb{R}^6$, $\tilde{\sigma} = [\sigma_1, \sigma_2, \ldots, \sigma_6] \in \mathbb{R}^6$ and $\tilde{\rho} = [\rho_1, \rho_2, \ldots, \rho_6] \in \mathbb{R}^6$. Sliding mode controller involves two phases – reaching phase and sliding phase. The sliding mode controller operates in two distinct phases, which include the reaching phase and the sliding phase to robustly control even in the presence of uncertainties and disturbances. In the reaching phase, the system states start from an arbitrary initial condition and move towards the sliding surface $\tilde{s} = 0$ as defined in Eq. 16 Once the system states reaches the sliding surface, the states remain constrained to the sliding surface in the sliding phase and converge to the equilibrium point. The reaching phase ensures the system reaches the sliding surface in a finite time, after which the sliding phase achieves precise trajectory tracking. However, reaching phase is sensitive to parametric variations and uncertainties. So, a fixed-time reaching law based control law is defined in Eq. 17 to ensure that the system states to reach the sliding surface within fixed time.

$$\dot{\tilde{s}} \quad = \quad -\mathrm{k}_{s1} sig^{l_1}(\tilde{s}) - k_{s2} sig^{l_2}(\tilde{s}) \tag{17}$$

**Theorem 1** *If the sliding surface is defined as in Eq. 16 for the relative translational and rotational dynamics described in Eq. 15, the configuration tracking error and relative velocity of the chaser satellite with respect to the target satellite will converge to zero (equilibrium point) along the sliding surface within a fixed time. The unified feedback control law for the coupled translational and rotational motion is as follows:*

$$\begin{aligned} \tilde{\varphi}_{c/c} \quad = \quad & -\tilde{\varphi}_{g/c} - \tilde{\varphi}_{d/c} - \mathbb{I}_c \Big( ad^*_{\tilde{\xi}_c} \mathbb{I}_c \tilde{\xi}_c + ad_{\tilde{\sigma}} Ad_{\boldsymbol{H}^{-1}} \tilde{\xi}_t - Ad_{\boldsymbol{H}^{-1}} \dot{\tilde{\xi}}_t \\ & + (\mathbb{Q}_1 + \mathbb{Q}_2) \boldsymbol{G}(\tilde{\rho}) \tilde{\sigma} \Big) - \mathrm{k}_{s1} sig^{l_1}(\tilde{s}) - k_{s2} sig^{l_2}(\tilde{s}) \end{aligned} \tag{18}$$

*where* $\mathbb{Q}_1 = diag(\mathrm{k}_1 l_1 |\tilde{\rho}|^{l_1 - 1})$ *and* $\mathbb{Q}_2 = diag(\mathrm{k}_2 l_2 |\tilde{\rho}|^{l_2 - 1})$.

*Proof* Consider a continuous differentiable Lyapunov function:

$$V(\tilde{\sigma}, \tilde{\xi}_c, \boldsymbol{H}, \tilde{\xi}_t) = \frac{1}{2} \tilde{s}^T \mathbb{I}_c \tilde{s}$$

The time derivative of the Lyapunov function yields, $\dot{V}(\tilde{\sigma}, \tilde{\xi}_c, \boldsymbol{H}, \tilde{\xi}_t) = \tilde{s}^T \mathbb{I}_c \dot{\tilde{s}}$

$$\begin{aligned} \dot{V}(\tilde{\sigma}, \tilde{\xi}_c, \boldsymbol{H}, \tilde{\xi}_t) \quad &= \quad \tilde{s}^T \mathbb{I}_c \big( \dot{\tilde{\sigma}} + \mathrm{k}_1 l_1 |\tilde{\rho}|^{l_1 - 1} \dot{\tilde{\rho}} + \mathrm{k}_2 l_2 |\tilde{\rho}|^{l_2 - 1} \dot{\tilde{\rho}} \big) \\ &= \quad \tilde{s}^T [\tilde{\varphi}_{g/c} + \tilde{\varphi}_{d/c} + \tilde{\varphi}_{c/c} + \mathbb{I}_c ( ad^*_{\tilde{\xi}_c} \mathbb{I}_c \tilde{\xi}_c + ad_{\tilde{\sigma}} Ad_{\boldsymbol{H}^{-1}} \tilde{\xi}_t \\ & \quad - Ad_{\boldsymbol{H}^{-1}} \dot{\tilde{\xi}}_t + \mathbb{Q}_1 \boldsymbol{G}(\tilde{\rho}) \tilde{\sigma} + \mathbb{Q}_2 \boldsymbol{G}(\tilde{\rho}) \tilde{\sigma})] \end{aligned} \tag{19}$$

Substituting the control law defined in Eq. 18 into Eq. 19 yields

$$\begin{aligned} \dot{V}(\tilde{\sigma}, \tilde{\xi}_c, \boldsymbol{H}, \tilde{\xi}_t) \quad &= \quad \tilde{s}^T (-\mathrm{k}_{s1} sig^{l_1}(\tilde{s}) - k_{s2} sig^{l_2}(\tilde{s})) \\ &= \quad -\sum_{i=1}^{6} \mathrm{k}_{s1,i} (|s_i|^2)^{\frac{l_1+1}{2}} - \sum_{i=1}^{6} \mathrm{k}_{s2,i} (|s_i|^2)^{\frac{l_2+1}{2}} \end{aligned}$$

where $0 < \frac{l_1+1}{2} \leq 1$ and $\left(\frac{l_2+1}{2}\right) > 1$. So, according to **Lemma 2**, the derivative of Lyapunov function can be written as shown below.

$$\begin{aligned} \dot{V}(\tilde{\sigma}, \tilde{\xi}_c, \boldsymbol{H}, \tilde{\xi}_t) \quad &\leq \quad -\mathrm{k}_{s1} \left( \sum_{i=1}^{6} |s_i|^2 \right)^{\frac{l_1+1}{2}} - \mathrm{k}_{s2} 6^{\frac{l_2-1}{2}} \left( \sum_{i=1}^{6} |s_i|^2 \right)^{\frac{l_2+1}{2}} \\ \dot{V}(\tilde{\sigma}, \tilde{\xi}_c, \boldsymbol{H}, \tilde{\xi}_t) \quad &\leq \quad -k_{t1} V^{\frac{l_1+1}{2}} - k_{t2} V^{\frac{l_2+1}{2}} \end{aligned}$$

where $\lambda_{min}(\mathrm{k}_{s1})$ *and* $\lambda_{min}(\mathrm{k}_{s2})$ are minimum eigen value of $k_{s1}$ and $k_{s2}$, respectively, $k_{t1} = \left(2^{\frac{l_1+1}{2}}\right) \lambda_{min}(\mathrm{k}_{s1})$ and $k_{t2} = \left(2^{\frac{l_1+1}{2}}\right)\left(6^{\frac{l_2-1}{2}}\right) * \lambda_{min}(\mathrm{k}_{s2})$. Therefore, by Lemma 1, the system states will reach the desired sliding surface $\tilde{s} = 0$ given by Eq. 16 within fixed-time, $T_{max}$ as shown below.

$$T(x) \leq T_{max} = \frac{2}{k_{t1}(1 - l_1)} + \frac{2}{k_{t2}(l_2 - 1)}$$

The closed loop feedback system described by Eq. 15 is globally fixed-time stable based on the control input given by Eq. 18 The configuration tracking error $\tilde{\rho}$ and the relative velocity $\tilde{\sigma}$ of chaser satellite with respect to the target satellite converges to zero within fixed-time.

## 3.2 Noise Transformation in Manifold

We have added noise to the translational and rotational states of the system to compare how the proposed adaptive gain tuning-based fixed-time sliding mode controller perform with respect to the conventional fixed-time sliding mode controller. The noise is modeled using a manifold-centered multi-variate Gaussian distribution approach that separates the mean on the manifold and covariance in the tangent space. Lie groups are nonlinear manifolds where the standard vector-space operations are undefined. The tangent space provides a local linear approximation where Gaussian noise can be defined without violating the manifold's constraints. The nominal state $\tilde{\rho} \in \mathbb{R}^6$ representing translational and orientation states acts as the mean, which evolves on the $SE(3)$ manifold. The noise vector $\tilde{\mathbb{m}} = [\tilde{\mathbb{w}}, \tilde{\mathbb{v}}]^T \in \mathbb{R}^6$ encodes translational and rotational perturbations in the vector space and is given by $\tilde{\mathbb{m}} \sim \mathcal{N}(0, \Sigma)$. The covariance matrix $\Sigma \in \mathbb{R}^{6\times6}$ is a positive-semidefinite matrix defined in the tangent space at the identity of Lie group and is shown below.

$$\Sigma = \begin{bmatrix} \Sigma_{\mathbb{w}} & \Sigma_{\mathbb{w}\mathbb{v}} \\ \Sigma_{\mathbb{v}\mathbb{w}} & \Sigma_{\mathbb{v}} \end{bmatrix} \in \mathbb{R}^{6\times6}$$

where $\Sigma_{\mathbb{w}}$ and $\Sigma_{\mathbb{v}}$ represents independent noise about rotational and translational part, respectively, and $\Sigma_{\mathbb{w}\mathbb{v}}$ represents cross-covariance noise between rotation and translational. The exponential mapping of the noise vector $\tilde{\mathbb{m}}$ is performed to propagate the perturbations from tangent space to the manifold. Then the noisy state vector is obtained via group multiplication as $\tilde{\rho}_{noisy} = \exp(\tilde{\mathbb{m}})\tilde{\rho}$. The covariance matrix used in this paper is as shown below

$$\Sigma = \begin{bmatrix} 0.01\boldsymbol{I}_3 & 0_{3\times3} \\ 0_{3\times3} & 0.05\boldsymbol{I}_3 \end{bmatrix} \in \mathbb{R}^{6\times6}$$

## 3.3 Reinforcement Learning based Neural Network

The reaching law described in Eq. 17 ensures error states reach the sliding surface within fixed time. Reaching law is dependent on two constants $k_{s1}$ and $k_{s2}$ and plays a crucial role in determining the system's ability to achieve and maintain stability in the presence of uncertainties and parameter variations. These gains directly influence how quickly and effectively the system can transition onto the sliding surface. The fixed time control law for reaching phase can be written as

$$\tilde{\varphi}_s(t) = -k_{s2} sig^{l_1}(\tilde{s}) - k_{s2} sig^{l_2}(\tilde{s}) \tag{20}$$

By discretizing the continuous switching control input as shown in Eq. 20

$$\begin{aligned} \Delta\tilde{\varphi}_s(t) &= \tilde{\varphi}_s(t) - \tilde{\varphi}_s(t-1) \\ &= -k_{s1} sign(s)(|s(t)|^{l_1} - |s(t-1)|^{l_1}) - k_{s2} sign(s)(|s(t)|^{l_2} - |s(t-1)|^{l_2}) \end{aligned} \tag{21}$$

The proposed adaptive controller is a cascade-based architecture as shown in Fig. 1 below. The sliding mode controller is placed in cascade with the plant neural network which tunes the gains of reaching law $k_{s1}$ and $k_{s2}$. The gains consists of static and dynamic part. The static part is the initial gain value and the variable part $\Delta k_{si}$ is tuned by the neural network as shown below.

$$k_{si} = k_{si} + \Delta k_{si} \quad , i = 1,2$$

The neural network takes input as configuration errors $(\tilde{\rho}(t), \tilde{\rho}(t-1))$, sliding variable $(\tilde{s}(t), \tilde{s}(t-1))$ and reaching phase control input $(\tilde{\varphi}_s(t), \tilde{\varphi}_s(t-1))$. The neural model for the system can be expressed as

$$\Delta k_{si}(t+1) = f(\tilde{\rho}(t), \tilde{\rho}(t-1), \tilde{s}(t), \tilde{s}(t-1), \tilde{\varphi}_s(t), \tilde{\varphi}_s(t-1)) \quad , i = 1,2$$

We have employed a hybrid offline-online based neural network learning approach to tune the gains of the sliding mode controller in real time. The neural network is trained offline using dataset obtained from two different rendezvous and docking maneuver scenario based on conventional fixed-time sliding mode controller. The backpropagation algorithm is used to adjust the weights and biases of the network during offline training. The neural network uses the pre-trained weights and biases to estimate the gains in real time. Thus, eliminating computationally expensive training during real-time operation and enabling faster response time. Sigmoid activation function is used in both the layers.

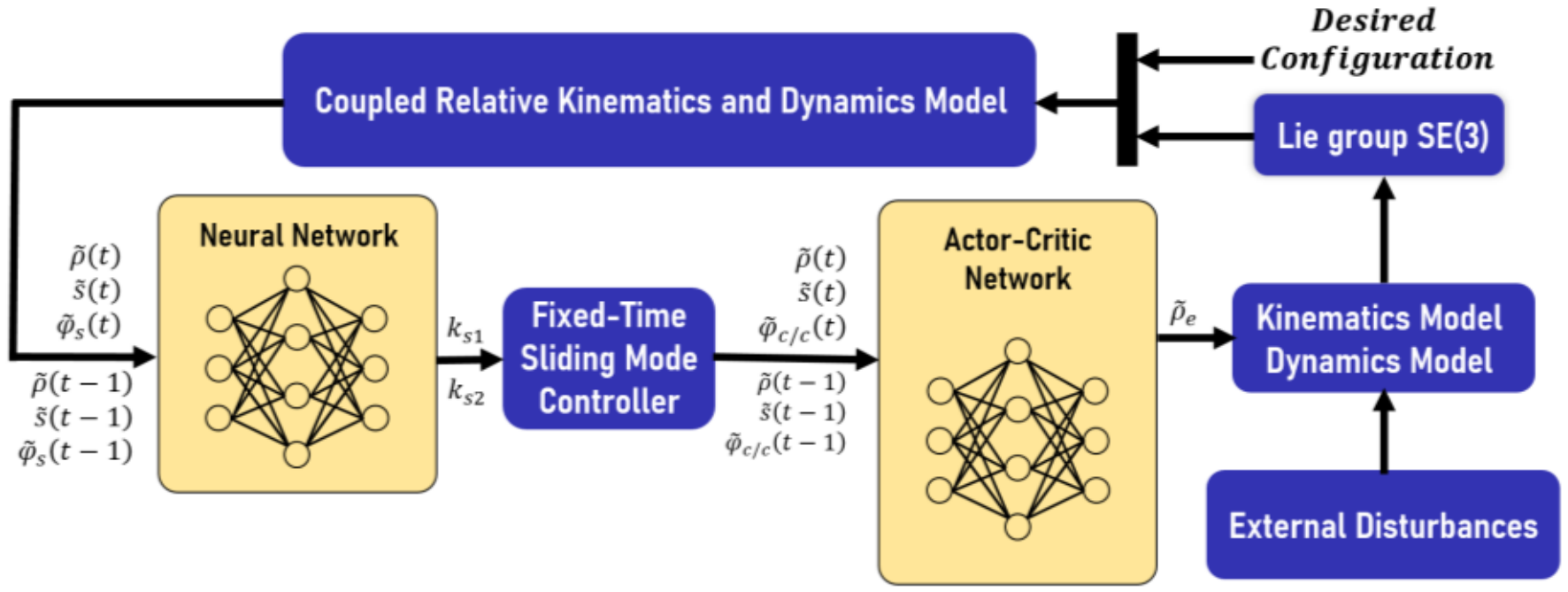

*Figure 1 Hybrid control architecture of neural network based Sliding Mode Controller*

The gains of the sliding mode controller are adjusted to minimize the cost function given by

$$J(t) = \frac{1}{2}\left(\tilde{\rho}(t) - \tilde{\rho}_e(t)\right)^2$$

where $\tilde{\rho}_e$ is the estimated value of configuration tracking error. The variable component of the gains of the reaching law are tuned in each cycle. In order to make the learning rate faster the exponentially weighted moving average of the gain update of past iterations are considered. Using this approach if $\Delta k_{s1}$ for recent past iterations are high then the current change will be high so to converge faster and vice versa. The updated gain law can be formulated as follows.

$$\Delta k_{s1} = -\alpha\left(\frac{\partial J(t)}{\partial k_{s1}}\right) + \gamma \Delta k_{s1}(t-1) + \gamma(1-\gamma)\Delta k_{s1}(t-2)$$

where, $\alpha$ is the learning rate and $\gamma$ is the momentum rate

$$\frac{\partial J(t)}{\partial k_{s1}} = \frac{\partial J(t)}{\partial \tilde{\rho}_e} * \frac{\partial \tilde{\rho}_e}{\partial \tilde{\varphi}_s} * \frac{\partial \tilde{\varphi}_s}{\partial k_{s1}}$$

$$= -(\tilde{\rho} - \tilde{\rho}_e) * \left(\tilde{\rho}_e(1 - \tilde{\rho}_e) \sum_j W_{i,h} Y_{h,j}(1 - Y_{o,j}) W_{h,o}\right) * \left(\frac{\partial \tilde{\varphi}_s}{\partial k_{s1}}\right)$$

where $\frac{\partial \tilde{\varphi}_s}{\partial k_{s1}} = -sign(s)(|s(t)|^{l_1} - |s(t-1)|^{l_1})$

$Y_{h,j}$ and $Y_{o,j}$ is the output of j^th^ neuron of hidden layer and output layer, respectively. $W_{i,h}$ and $W_{h,o}$ are weights of the neural model from the input layer to hidden layer, and from hidden layer to output layer, respectively. Similarly, it can be computed for the gain $k_{s2}$. The system control input obtained through gain tuning network $\left(\tilde{\varphi}_c(t), \tilde{\varphi}_c(t-1)\right)$ is fed to the system identification network. This network estimates the new output of the system using Actor-Critic network architecture. Actor based neural network is used to estimates the output of the system based on mean and variance whereas the critic based network evaluates the actions taken by the actor using value function and using this approach gives feedback to the actor to perform better. The actor network gives output as the estimated configuration error $\tilde{\rho}_e$ as shown below by minimizing the estimation error $(\tilde{\rho}_e - \tilde{\rho})$.

$$\tilde{\rho}_e(t+1) = f(\tilde{\rho}(t), \tilde{\rho}(t-1), \tilde{s}(t), \tilde{s}(t-1), \tilde{\varphi}_s(t), \tilde{\varphi}_s(t-1))$$

The critic network minimizes the value function error, $\vartheta(t) - \vartheta(t-1)$ and reward function given by $\mathcal{R}(t) = -r_1(\tilde{\rho}_e(t) - \tilde{\rho}(t))^2 - r_2 \tilde{\varphi}_{c/c}(t)$ which are together known as the temporal difference error $\varepsilon_{TDE}$

$$\varepsilon_{TDE}(t) = \mathcal{R}(t) + \varrho(\vartheta(t) - \vartheta(t-1))$$

where $0 < \varrho \leq 1$ is discount factor describing the decay of value function error. The value function denoted by $\vartheta(t)$ is defined as

$$\vartheta(t) = \sum_{j=1}^{k} W_{h,0} Y_{h,j}$$

where j represents the $j^{th}$ neuron of hidden layer; $W_{h,0}$ represents weights between hidden layer and output layer; and $Y_{h,j}$ represents output of $j^{th}$ neuron of hidden layer.

# 4 Results and Discussion

Target spacecraft is assumed to be in Molniya orbit whose eccentricity is 0.72 and the value of other classical orbital elements of the orbit is given in Table 1. The mass of target spacecraft is taken as 110 kg while the mass of the chaser spacecraft is taken as 100 kg. The inertia matrix of chaser and target spacecraft is represented by $\boldsymbol{J}_c$ and $\boldsymbol{J}_t$ respectively as

$$\boldsymbol{J}_c = \begin{bmatrix} 50 & 0 & 0 \\ 0 & 100 & 0 \\ 0 & 0 & 50 \end{bmatrix} \quad and \quad \boldsymbol{J}_t = \begin{bmatrix} 150 & 0 & 0 \\ 0 & 50 & 0 \\ 0 & 0 & 50 \end{bmatrix}$$

The results of the simulation were compared as shown in Fig. 2-5 where the plots shown in left column (a) are obtained by using conventional fixed-time Sliding Mode Controller (SMC) whereas the plots shown in the right column (b) is obtained using neural network based adaptive fixed-time sliding mode controller. It can be inferred from the plots of the relative position and attitude tracking error as shown in Fig. 2 and Fig. 3, respectively that the convergence rate of the proposed controller is faster than the convention fixed-time SMC.

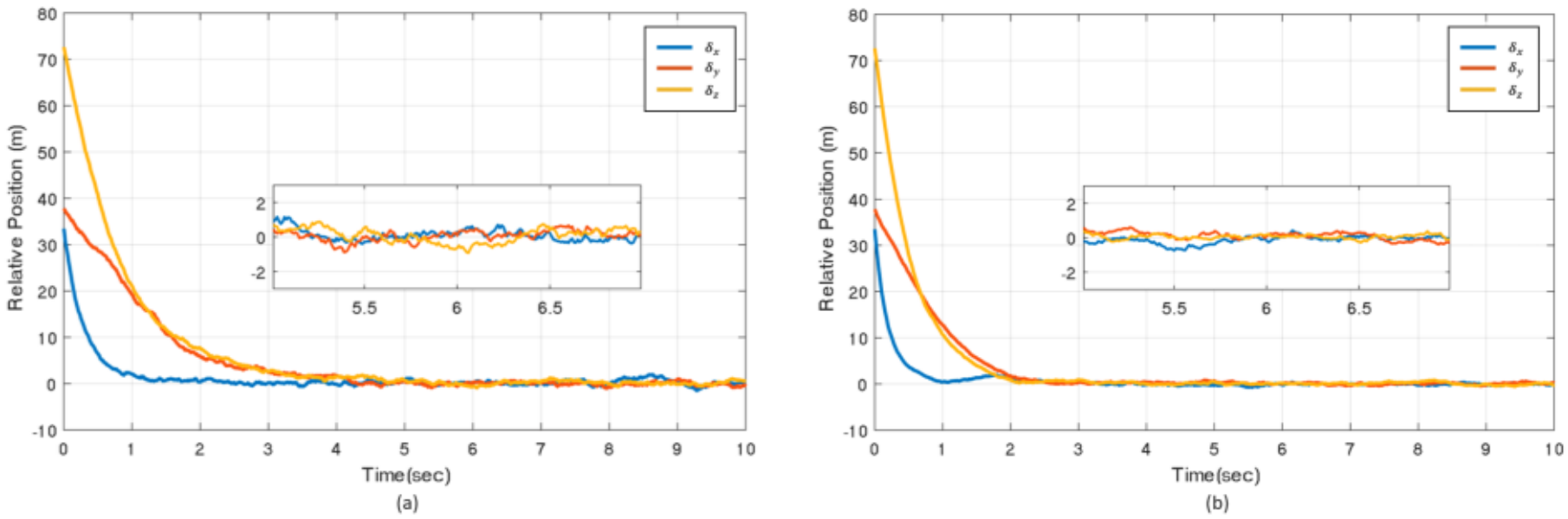

*Figure 2 Relative Position (a) Fixed-Time SMC (b) Neural Network based SMC*

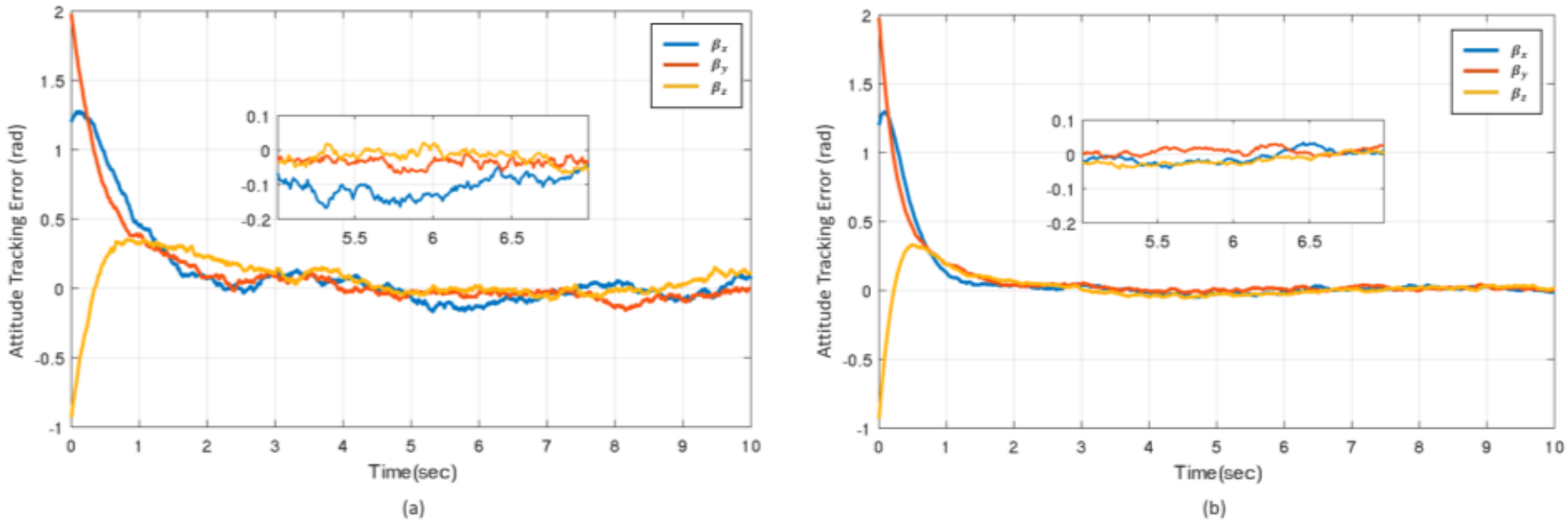

*Figure 3 Attitude Tracking Error (a) Fixed-Time SMC (b) Neural Network based SMC*

Moreover, the plot of tracking error shows that using neural network based fixed-time SMC the reaching phase is quite smaller than the one with conventional fixed-time SMC. The sliding variable converges within 4 second in the proposed adaptive neural network based fixed-time sliding mode controller. However, in conventional fixed-time SMC there is still chattering or oscillation about origin even after reaching the sliding surface.

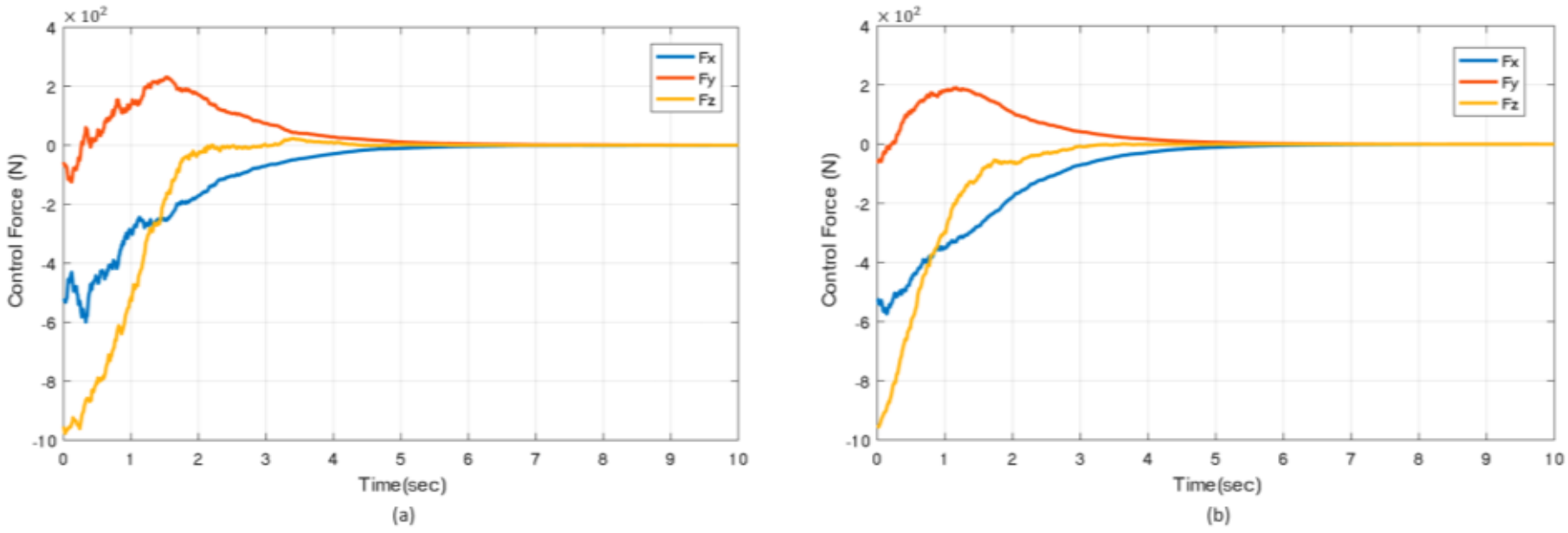

*Figure 4 Control Input Force (a) Fixed-Time SMC (b) Neural Network based SMC*

The plot of control input force and torque as shown in Fig. 4 and Fig. 5, respectively shows that using neural network based fixed-time SMC the control input is relatively smoother than the one with conventional fixed-time SMC.

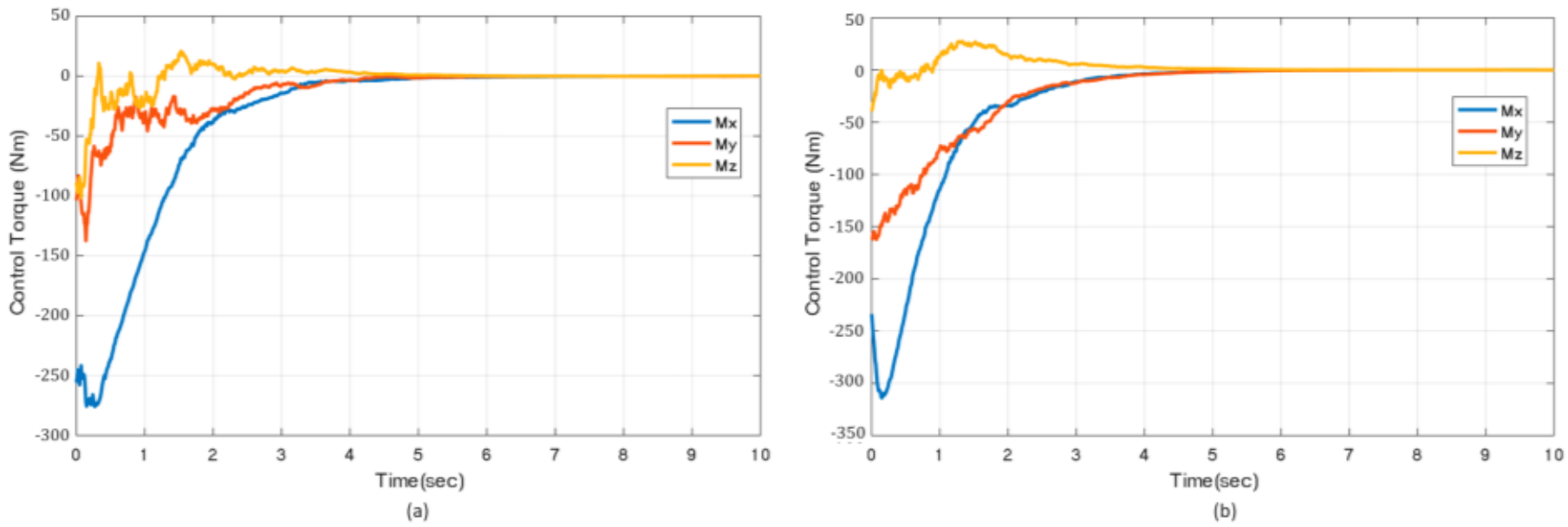

*Figure 5 Control Input Torque (a) Fixed-Time SMC (b) Neural Network based SMC*

As we have tuned the gains of the sliding surface based on the trained weights and biases of the network in offline mode, the error does not converge to zero as desired. However, the results obtained using neural network are better than the conventional SMC. In future we can train the network weights and biases in online mode to improve the results even better.

# 5 Conclusion

In this paper, we have tuned the gains of the Fixed-time Sliding surface adaptively using neural network. We have assumed that we don't have an accurate model of the system so we have used reinforcement learning to estimate the dynamical model of the system. Then the estimated model is used to estimate the states of system using the Actor-Critic method. The proposed control algorithm integrates a neural network and a sliding mode controller in a cascade loop architecture, where the tracking error dynamically tunes the sliding surface slope.

# References

1. Li Q, Yuan J, Zhang B, Gao C (2017) Model predictive control for autonomous rendezvous and docking with a tumbling target. Aerospace Science and Technology; Volume 69:700–711. https://doi.org/10.1016/j.ast.2017.07.022
2. Arantes G, Martins-Filho LS (2014) Guidance and Control of Position and Attitude for Rendezvous and Dock/Berthing with a Noncooperative/Target Spacecraft. Mathematical Problems in Engineering; Volume 2014, Issue 1:508516. https://doi.org/10.1155/2014/508516
3. Polites ME (2012) Technology of Automated Rendezvous and Capture in Space. Journal of Spacecraft and Rockets; Volume 36, Issue 2. https://doi.org/10.2514/2.3443
4. Fehse W (2003) Automated Rendezvous and Docking of Spacecraft. Cambridge University Press. https://doi.org/10.1017/cbo9780511543388
5. Boyarko G, Yakimenko O, Romano M (2012) Optimal Rendezvous Trajectories of a Controlled Spacecraft and a Tumbling Object. Journal of Guidance, Control, and Dynamics; Volume 34, Issue 4:1239–1252. https://doi.org/10.2514/1.47645
6. Kasiri A, Fani Saberi F (2023) Coupled position and attitude control of a servicer spacecraft in rendezvous with an orbiting target. Scientific Reports; 13:4182. https://doi.org/10.1038/s41598-023-30687-9

7. Yan H, Tan S, Xie Y (2018) Integrated Translational and Rotational Control for the Final Approach Phase of Rendezvous and Docking. Asian Journal of Control; Volume 20, Issue 5:1967–1978. https://doi.org/10.1002/asjc.1712
8. Yamanaka K, Ankersen F (2012) New State Transition Matrix for Relative Motion on an Arbitrary Elliptical Orbit. Journal of Guidance, Control, and Dynamics; Volume 25, Issue 1. https://doi.org/10.2514/2.4875
9. Song Y, Zhou Q, Chen Q (2024) Coupled translation and attitude tracking control for multi-satellite electromagnetic formation flight based on dual quaternion. Advances in Space Research; Volume 73, Issue 10:5299–5317. https://doi.org/10.1016/j.asr.2024.02.058
10. Zhang F, Duan G (2013) Robust adaptive integrated translation and rotation control of a rigid spacecraft with control saturation and actuator misalignment. Acta Astronautica; Volume 86:167–187. https://doi.org/10.1016/j.actaastro.2013.01.010
11. Lee D (2015) Spacecraft Coupled Tracking Maneuver Using Sliding Mode Control with Input Saturation. Journal of Aerospace Engineering; Volume 28, Issue 5. https://doi.org/10.1061/(asce)as.1943-5525.0000473
12. Gong K, Liao Y, Wang Y (2020) Adaptive Fixed-Time Terminal Sliding Mode Control on SE(3) for Coupled Spacecraft Tracking Maneuver. International Journal of Aerospace Engineering; Volume 2020, Issue 1:3085495. https://doi.org/10.1155/2020/3085495
13. Mortari D, Angelucci M, Markey F. L (2000) Singularity and attitude estimation. Spaceflight Mechanics:479-493
14. Bhat SP, Bernstein DS (1998) A topological obstruction to global asymptotic stabilization of rotational motion and the unwinding phenomenon. Systems & Control Letters,39:63-70. https://doi.org/10.1016/s0167-6911(99)00090-0
15. Mayhew CG, Sanfelice RG, Teel AR (2011) On quaternion-based attitude control and the unwinding phenomenon. Proceedings of the American Control Conference:299–304. https://doi.org/10.1109/acc.2011.5991127
16. Das G, Sinha M (2022) Unwinding-Free Fast Finite-Time Sliding Mode Satellite Attitude Tracking Control. Journal of Guidance, Control, and Dynamics, Volume 46, Issue 2:325–342. https://doi.org/10.2514/1.G006949
17. Lee T, Leok M, McClamroch NH (2010) Geometric tracking control of a quadrotor UAV on SE(3). Proceedings of the IEEE Conference on Decision and Control:5420–5425. https://doi.org/10.1109/CDC.2010.5717652
18. Shtessel Y, Edwards C, Fridman L, Levant A (2014) Sliding mode control and observation. https://doi.org/10.1007/978-0-8176-4893-0
19. Latosiński P (2017) Sliding mode control based on the reaching law approach - A brief survey. 22nd International Conference on Methods and Models in Automation and Robotics:519–524. https://doi.org/10.1109/mmar.2017.8046882
20. Wu Y. D, Wu S. F, Gong D. R, et al (2020) Spacecraft Attitude Maneuver Using Fast Terminal Sliding Mode Control Based on Variable Exponential Reaching Law. Lecture Notes in Electrical Engineering; 622. https://doi.org/10.1007/978-981-15-1773-0_1
21. Gong K, Liao Y, Wang Y (2020) Adaptive Fixed-Time Terminal Sliding Mode Control on SE(3) for Coupled Spacecraft Tracking Maneuver. International Journal of Aerospace Engineering; Volume 2020, Issue 1:3085495. https://doi.org/10.1155/2020/3085495

22. Tokat S, Eksin I, Guzelkaya M (2009) Linear time-varying sliding surface design based on co-ordinate transformation for high-order systems. Transactions of the Institute of Measurement and Control; Volume 31, Issue 1. https://doi.org/10.1177/0142331208090672
23. Salamci MU, Tombul GS (2006) Sliding mode control design with time varying sliding surfaces for a class of nonlinear systems. Proceedings of the IEEE International Conference on Control Applications:996–1001. https://doi.org/10.1109/cacsd-cca-isic.2006.4776780
24. Sahoo RK, Sinha M (2024) Coupled Rendezvous and Docking Maneuver Control of Spacecraft using Fast Fixed-Time Sliding Mode Controller. IAF Astrodynamics Symposium; Volume 1:1028-1042. https://doi.org/10.52202/078368-0087
25. Humaidi AJ, Hasan AF (2019) Particle swarm optimization–based adaptive super-twisting sliding mode control design for 2-degree-of-freedom helicopter. Measurement and Control (United Kingdom); Volume 52, Issue 9-10:1403–1419. https://doi.org/10.1177/0020294019866863
26. Rajendran S, Muthu Kumar A, Agnes Idhaya Selvi V (2022) Particle Swarm Optimization tuned Hybrid Sliding Mode Controller based static synchronous compensator with LCL filter for Power Quality improvement. Sustainable Energy Technologies and Assessments; Volume 53, Part C:102653. https://doi.org/10.1016/J.SETA.2022.102653
27. Hollweg GV, de Oliveira Evald PJD, Mattos E, et al (2023) Self-tuning methodology for adaptive controllers based on genetic algorithms applied for grid-tied power converters. Control Engineering Practice; Volume 135:105500. https://doi.org/10.1016/j.conengprac.2023.105500
28. Teklu EA, Abdissa CM (2023) Genetic Algorithm Tuned Super Twisting Sliding Mode Controller for Suspension of Maglev Train With Flexible Track.Volume 11:30955–30969. https://doi.org/10.1109/access.2023.3262416
29. Yang Y, Wang Y, Zhang W, et al (2022) Design of Adaptive Fuzzy Sliding-Mode Control for High-Performance Islanded Inverter in Micro-Grid. Energies; Volume 15, Issue 23:9154. https://doi.org/10.3390/en15239154
30. Gu C, Chi E, Guo C, et al (2023) A New Self-Tuning Deep Neuro-Sliding Mode Control for Multi-Machine Power System Stabilizer. Mathematics; Volume 11, Issue 7:1616. https://doi.org/10.3390/math11071616
31. Truong H, Tran D, et al (2018) A Neural Network Based Sliding Mode Control for Tracking Performance with Parameters Variation of a 3-DOF Manipulator. Applied Sciences, Volume 9, Issue 10. https://doi.org/10.3390/app9102023
32. Massou S, Boumhidi I (2023) Adaptive control based neural network sliding mode approach for two links robot. International Journal of Power Electronics and Drive Systems (IJPEDS); Volume 14:2546–2556. https://doi.org/10.11591/ijpeds.v14.i4.pp2546-2556
33. Bullo, F., & Murray, R.M. (1995). Proportional Derivative (PD) Control on the Euclidean Group
34. Zuo, Z. (2015). Non-singular fixed-time terminal sliding mode control of non-linear systems. IET Control Theory & Applications; Volume 9, Issue 4:545-552. https://doi.org/10.1049/iet-cta.2014.0202